# Femtosecond low-threshold all-optical switching enabled by giant broadband optical nonlinearity from heteroatom doping

Xuefeng Zhang[1, #], Zhongquan Nie[1, #,*], Jinhai Zou[1], Hanke Zhang[1], Xiaobin Wang[1], Jian Wu[1], Yuqiu Zhang[1], Jinyong Leng[1], Guohui Li[2], Yue Tian[3], Baohua Jia[4,*], Pu Zhou[1, *]

[1] College of Advanced Interdisciplinary Studies, National University of Defense Technology, Changsha, 410073, China

[2] College of Physics and Photo Electricity Engineering, Key Lab of Advanced Transducers and Intelligent Control System of Ministry of Education, Taiyuan University of Technology, Taiyuan, 030024 China

[3] School of Chemistry and Chemical Engineering, Hainan University, Haikou 570228, China

[4] Centre for Atomaterials and Nanomanufacturing (CAN), School of Science, RMIT University, Melbourne, 3000 VIC, Australia

*Corresponding author: E-mail: niezhongquan1018@163.com (Z. Nie), baohua.jia@rmit.edu.au (B. Jia), zhoupu203@163.com

[#]These authors contributed equally to this paper

## Abstract

Ultrafast all-optical switching (AOS) is pivotal for advancing integrated photonic devices, from high-speed photonic information processing to next generation all-optical computing and communication networks. However, conventional nonlinear materials suffer from sluggish response time, high power threshold, weak and narrow-bandwidth optical nonlinearities, critically limiting their viability. Here, we report a heteroatom engineering strategy to overcome these limitations by designing zero-dimensional nitrogen-doped carbon quantum dots (N-CQDs) with nonlinear optical performance far exceeding the state-of-the-art. Leveraging spatial self-phase modulation (SSPM) and ultrafast pump-probe technique, we first demonstrate an all-in-one AOS platform, where femtosecond laser pulses serve dual roles as control and signal beams. The AOS simultaneously realizes ultrafast response time (~520 $fs$), ultralow threshold energy (2.2 Wcm$^{-2}$), and giant nonlinear refraction indexes (~10$^{-5}$ cm$^2$/W) in the wide spectral range (400 ~ 1064 nm), yielding performance surpassing state-of-the-art nonlinear carbon materials (i.e.

carbon nanotube) by orders of magnitude. Spectroscopic and bandgap analyses attribute these exotic performances to enhanced n-$\pi^*$ interaction enabled by nitrogen doping, which amplifies nonlinear polarization dynamics. Crucially, ultrafast fluorescence spectroscopy reveals a large two-photon absorption cross-section of the N-CQDs, challenging the conventional cognition that broadband SSPM necessitates single-photon excitation. This discovery unveils a multi-channel AOS rooted in synergistic single- and two-photon process. This work demonstrates a new paradigm for achieving ultrafast, broadband, and energy-efficient AOS by heteroatom doping engineering. Their exceptional performance positions them as an emerging disruptive nonlinear material for applications in AOS, optical logic gates, and nonlinear signal processing, heralding new frontiers in photonic device innovation.

***Key words***: Ultrafast pulse laser, Optical nonlinearity, Self-phase modulation, Carbon quantum dots, All-optical switching

# 1. Introduction

Ultrafast all-optical switching (AOS) capable of controlling light through light is the fundamental building block in integrated photonic systems, including photonic information processing, all-optical network communication and high-performance optical computing.[1-3] Unlike electronic switching,[4] AOS, particularly when tailored by nonlinear optical spatial self-phase modulation (SSPM), holds huge potential to surpass electronic performance limits in speed and energy efficiency. However, critical challenges persist in existing nonlinearity material driven AOS: (i) Slow response time on the order of several hundred picoseconds[1, 5] to even nanosecond,[6] which hinders ultrafast light manipulation. (ii) Excessive energy thresholds on the order of several GWcm$^{-2}$ [7, 8] that impede cost-effective integration into photonic platforms[9] and induce thermal degradation risks. (iii) Weak, narrowband nonlinearities that restrict SSPM efficiency and limit multi-channel, onmi-directional AOS applications.[10] Addressing these intertwined limitations demands the development of advanced nonlinear optical materials capable of unifying ultrafast switching (up to femtosecond scale), ultralow power operation (down to Wcm$^{-2}$ level), and giant broadband nonlinearity (refraction index >10$^{-6}$ cm$^2$/W and waveband >200 nm), which is expected to open the door to novel applications in quantum computing and all-optical communication.

In this regard, a plethora of representative optical nonlinear materials, embracing metal nanomaterials,[11, 12] epsilon-near-zero (ENZ) materials[13-15] and two-dimensional (2D) materials,[16, 17] have gained a surge of interest in developing versatile AOS prototypes. Among them, metal nanocomposites/microstructures, while benefiting from the inherent quantum confinement-induced ultrafast modulation speed, exhibit weak optical nonlinearity in the non-resonant absorption windows and suffer from high energy consumption due to low excitation efficiency and large transmission loss[18] in the vicinity of resonant absorption regions. Alternatively, ENZ materials emerge as a promising AOS solution due to their large optical nonlinearities, which stem from the massive near-field enhancement at specific wavelengths. However, they only work at narrow waveband[19, 20] accompanying with relatively slow switching speed. By contrast, 2D layered materials have demonstrated extraordinary potential in SSPM-based multifunctional AOS owing to their wide spectral responses, large optical nonlinearities and fast electronic migration rates.[21-24] but their practical adoption is hindered by complex synthesis, low absorption in ultrathin layers, and high power demands.[25] Despite extensive efforts and significant progresses in AOS, the ability to fulfill the compatibility of aforementioned requirements remains to be elusive in the existent material systems. Therefore, how to engineer and develop novel nonlinear optical materials to realize the three-in-one (ultrafast switching, high efficiency, and giant and broadband optical nonlinearity) AOS is still in urgent demand.

Motivated by these challenges, here we first propose and demonstrate an ultrafast low-threshold and high-efficiency AOS from zero-dimensional (0D) nitrogen-doped carbon quantum dots (N-CQDs), enabled by ultrafast carrier dynamics and broadband nonlinear-based SSPM effect. Leveraging a dual-beam femtosecond laser system, we demonstrate light-controlled-light modulation in N-CQDs, enabled by strategic amino-group functionalization that optimizes donor-acceptor interactions. Spectroscopic and bandgap calculations validate the enhanced n-π* interaction in the π-conjugate system, thus facilitating exciton dissociation, broadband spectral absorption and accelerated charge transport. The N-CQDs, synthesized via a scalable hydrothermal route, exhibit ultrafast carrier dynamics in wavelength range from 532 to 740 nm on femtosecond-to-picosecond timescales, as validated by transient absorption spectroscopy. The ultrafast SSPM of the N-CQDs is further implemented to exhibit broadband nonlinear response at 400 ~ 1064 nm waveband. The nonlinear refractive index is determined to be as large as ~$10^{-5}$ $cm^2$/W by taking advantage of the linear relationship between SSPM ring number and excitation intensity. More interestingly, by jointly tuning both the excitation wavelength and repetition frequency of incident femtosecond laser pulse, we can boost the refractive indexes of the N-CQDs by two orders of magnitude, which is beneficial to scalable and ring-adjustable SSPM and high-efficiency AOS. Meanwhile, we also find that the excitation threshold of the AOS is as low as 2.2 $Wcm^{-2}$ according to the phase difference between the adjacent rings of the SSPM, which is three orders of magnitude lower than the state-of-the-art.[26] Beyond that, the ultrafast fluorescence spectroscopy reveals that the N-CQDs possess wider spectral response (400 ~ 1064 nm) and larger two-photon absorption (TPA) cross sections than the pristine CQDs, which confirms the physical origins of enhanced optical nonlinearity and SSPM-based high-performance AOS.

## 2. Heteroatom doping engineering and electronic structure insights

To shed light on the origin of the broadband large optical nonlinearities and ultrafast nonlinear dynamics of N-CQDs at the microscopic level, it is of pivotal significance to elucidate their electronic and band structures. Initially, it is assumed that the pristine CQDs consist of isolated $sp^2$ clusters embedded within the $sp^3$ carbon matrix, incorporating defects such as vacancies and edge-substituted surface functional groups (see Figure S1). A representative structure model is extracted from a partial graphene fragment on the $sp^2$ cluster rings, in which the CQDs with and without N-related intervention are constructed from 37 aromatic rings (see Figures 1a1, 1b1 and 1c1).

To further understand the electronic structure change in amine-functionalized CQDs, we evaluated the frontier molecular orbital energy levels, including the lowest unoccupied molecular orbital (LUMO) and the highest occupied molecular orbital (HOMO), using density functional theory (DFT) (see Figures 1a2, 1a3, 1b2, and 1b3, the computational details are provided in the Computational Methods section). This evaluation reveals the potential for ultrafast optical response, high electron mobility and wide-band absorption. It needs to mentioned that the HOMO and LUMO isosurfaces of pristine CQDs exhibit significant orbital overlap, whereas the relevant spatial separation appears clearly in the N-CQDs. These results suggest that the orbital of the lone pair electrons provided by amino groups can overlap with the conjugated π orbital, leading to the delocalization between the -$NH_2$ and CQDs conjugated system. On one hand, the reduced overlap leads to the formation of charge-separated excited states, whose rapid charge separation effectively suppresses the electron-hole recombination and enables ultrafast nonlinear response. On the other hand, the enhanced electron delocalization reinforces the coupling between the non-bonding (n) and π* orbitals, increasing transition probability and producing a more intense n-π* absorption band. Meanwhile, it is found that the 2p orbital of the N atom in amino groups aids in the formation of the HOMO level, as shown in Figure 1b2. The high charge density due to the existence of lone pairs of electrons in the p orbitals of the N atom makes -$NH_2$ prone to donate electrons. For N-CQDs, the anti-bonded p orbitals of conjugate system in Figure 1b3 participate in the formation of the LUMO level, thereby endowing the carbon atoms with electron-accepting properties. This implies that the intramolecular charge transfer process from the HOMO of -$NH_2$ to the LUMO of carbon structure occurs readily upon functionalizing amino groups, which can induce a new light absorption feature in the long-wavelength region.

Experimental absorption spectra in Figure 1a4 and 1b4 confirm that the CQDs functionalized with amine groups are capable of not only harvesting a wider absorption range compared to that of the pristine CQDs, but also maintaining large oscillator strengths of the fundamental radiative transitions. Besides, Pyrrole N and Graphite N as distinct N-related electronic modifications grafted on carbon skeleton also synergistically promotes the electron delocalization and spectral broadening, as shown in Figure 1c. The former localized on the carbon lattice contributes two p-electrons to the π-system, which primarily enhances the electron-donating capability (p-doping) of the carbon framework itself. The latter, on the contrary, creates an electron-deficient center and lowers the local energy, thus transforming the graphitic N site into a potent electron-accepting center. Additionally, the introduction of N in the CQDs decreases the HOMO−LUMO gap from 2.60 to 2.12 eV due to the charge transfer from three N configurations, as clearly depicted in Figure 1d.. Eventually, the charge density difference simulation further verifies the electron transfer from -NH$_2$ to carbon structure system, as presented in Figure 1e. The results show a yellow charge-depletion region around the N atom and a turquoise charge-accumulation region around the π-conjugated structure, from which the electron transfer number can be counted as 1.31. This efficient charge redistribution from -NH$_2$ to the whole carbon-structure plane, validates the enhanced charge delocalization and n-π* transition. In light of these theoretical insights, we demonstrate that amino-functionalized CQDs can achieve exceptional ultrafast response, efficient charge transport, and broad-spectrum optical absorption, enabling N-CQDs as highly promising and extraordinary nonlinear optical materials.

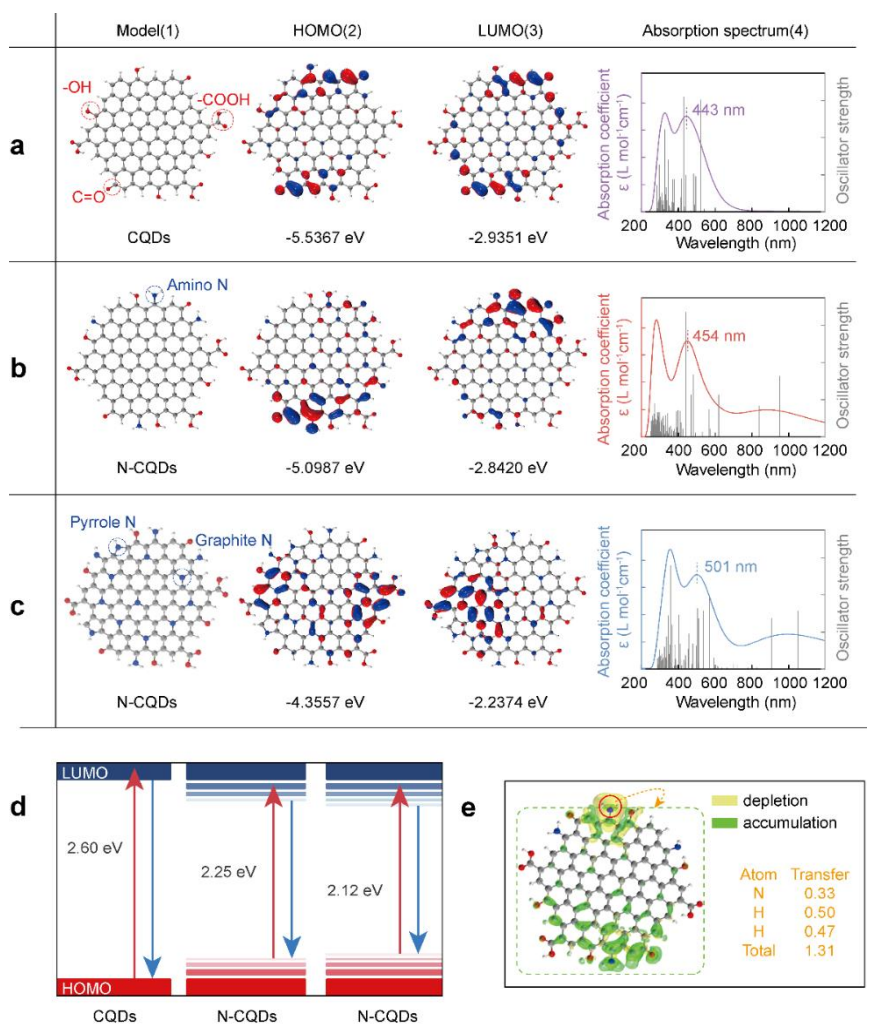

**Figure 1.** Time-dependent DFT calculation results. **a1** The proposed original CQDs molecular structure model together with several oxygen-containing functional groups. **b1** The amine group functionalized GQDs after structure optimization. **c1** The Pyrrole N and Graphite N functionalized GQDs after structure optimization. The calculated HOMO (**a2**, **b2**, **c2**), LUMO (**a3**, **b3**, **c3**), and absorption spectra (**a4**, **b4**, **c4**) of the CQDs and N-CQDs. **d** Relative energy level of the occupied and unoccupied molecular orbitals of the CQDs and N-CQDs; **e** The charge-density difference mapping of the N-CQDs.

## 3. Results and discussion

The N-CQDs were synthesized by the solvothermal strategy from L-Serine using acetic acid as the solvent (Figure 2a). Effective N-doping in CQDs is primarily governed by the critical carbonization and growth stages. The former enables the formation of N-containing heteroaromatic rings (N-6, N-5) within the graphitic core, while the latter facilitates the attachment of N-functional groups (e.g., -NH$_2$) to the surface via passivation. The synergistic effect of core doping during carbonization and surface functionalization during growth equips N-CQDs with excellent compatibility in electronic structure modulation and solution processability. Figure 2b displays a representative transmission electron microscopy (TEM) image of the N-CQDs, revealing uniformly distributed quasi-spherical nanodots. The corresponding high-resolution TEM image of an individual dot exhibits distinct lattice fringes of 0.22 nm, consistent with the (100) lattice spacing of graphene.[27] The particle size distribution of the N-CQDs yields an average diameter of 2.40 nm and sizes within a narrow range of 2.40 ± 0.28 nm. The wide-angle XRD pattern (Figure S2) of the N-CQDs shows an evident broad peak centered at around 22.6° corresponding to (002), suggesting partial graphitization and nanoscale crystalline feature of the N-CQDs.[28] Raman spectroscopy (Figure S3) shows characteristic D and G bands at 1341 cm$^{-1}$ (sp$^2$-hybridized D band) and 1595 (sp$^3$-hybridized G band).[29, 30] The intensity ratio (0.91) of the D to G band ($I_D/I_G$) indicates moderate amount of graphitic carbon inside the N-CQDs and contains high degree of structural defects. [31] These structural characterizations support the validity of our proposed theoretical model of N-CQDs.

Elemental analysis by energy-dispersive X-ray spectroscopy (EDS) confirms the presence of nitrogen doping within N-CQDs, with the atomic ratios of the N content determined at 15.5% (Figure 2c). Additionally, Fourier-transform infrared (FTIR) spectra (Fig. 2d) of the N-CQDs reveal the absorption bands at 3190, 1459 and 748 cm$^{-1}$, corresponding to N−H stretching, bending and waging vibrations, respectively, confirming the nitrogen content on the surface.[32] The wide-scan X-ray photoelectron spectroscopy (XPS) spectra indicate that the N-CQDs are composed of three elements (Figure S4), that is, C, O and N, corresponding to C 1s (284.5 eV), O 1s (532.6 eV) and N 1s (401.9 eV). High-resolution N 1s spectrum is presented in Figure 2e, where the three peaks at 399.2, 399.9, and 401.1 eV are assigned to pyrrolic N, Graphitic N, and amino N groups, respectively. Overall, these comprehensive analyses demonstrate the successful incorporation of nitrogen atoms into the CQD lattice and the consequent structural modification, which are anticipated to play a pivotal part in enhancing optical nonlinearity.

UV-vis absorption spectra of both the pristine CQDs and N-CQDs are shown in Fig. 1f. Pristine CQDs exhibit an absorption peak around 233 nm, corresponding to the π–π* transitions of the aromatic C=C bonds, and a shoulder at 332 nm due to the n–π* transition of C=O bond.[33, 34] However, its high optical transmittance beyond 600 nm limits its further nonlinear optical applications due to insufficient light-matter interaction in this spectral range. In contrast, the absorption spectrum of the N-CQDs shows a longer tail extending into the near-infrared range due to the N-induced electronic states,[28] enabling broad optical response and improved nonlinear performance. The inset shows the fluorescence behavior of a mercury lamp at 365 nm as the light excitation source. Photoluminescence (PL) and two-photon fluorescence properties of the N-CQDs were further systemically investigated with a custom-built fluorescence detection system Fluorescence detection system (Figure 1g). This setup, integrating a broadband femtosecond excitation source with a highly sensitive CCD spectrometer, allowed for the detection of weak optical signals with high sensitivity. As shown in Figure 1h, one-photon fluorescence excitation (green curve) at 370 nm yields a broadband PL emission peak at 467 nm, characteristic of blue fluorescence. Meanwhile, a representative two photon-induced PL spectrum of the N-CQDs is shown as a red curve with the 800 nm femtosecond pulsed laser excitation (Figure 1i). It can be found that its bandwidth is wider than that of the single photon one, and thus shows an almost white light emission at near 510 nm. Figure 1j illustrates the quadratic dependence of PL intensity on excitation power, with a fitting result showing that slope = 1.12, confirming the TPA nature of the N-CQD white light emission. Besides, the TPA cross-section of the N-CQDs is estimated using rhodamine 6G as the reference.[35] Before this, the quantum yields (QY) of the CQDs and N-CQDs are determined to be 25.7% and 31.7%, respectively, indicating a modest increase upon nitrogen doping.

The calculated TPA cross-section of N-CQDs reaches 47,000 GM (Equation S1), surpassing that of the undoped CQDs (34,000 GM) and even comparable to that of the CdTe quantum dots.[36] The large TPA cross-section of the N-CQDs may be ascribed to the existence of large π-conjugated systems and efficient intramolecular charge transfer enabled by nitrogen doping, which serves as a strong electron donor,[37, 38] enhancing the two-photon absorption and thus induing strong two-photon-induced fluorescence. In light of their broadband spectra response, high QY and large TPA cross-section, the N-CQDs hold a great promise for developing broad nonlinear photonics applications.

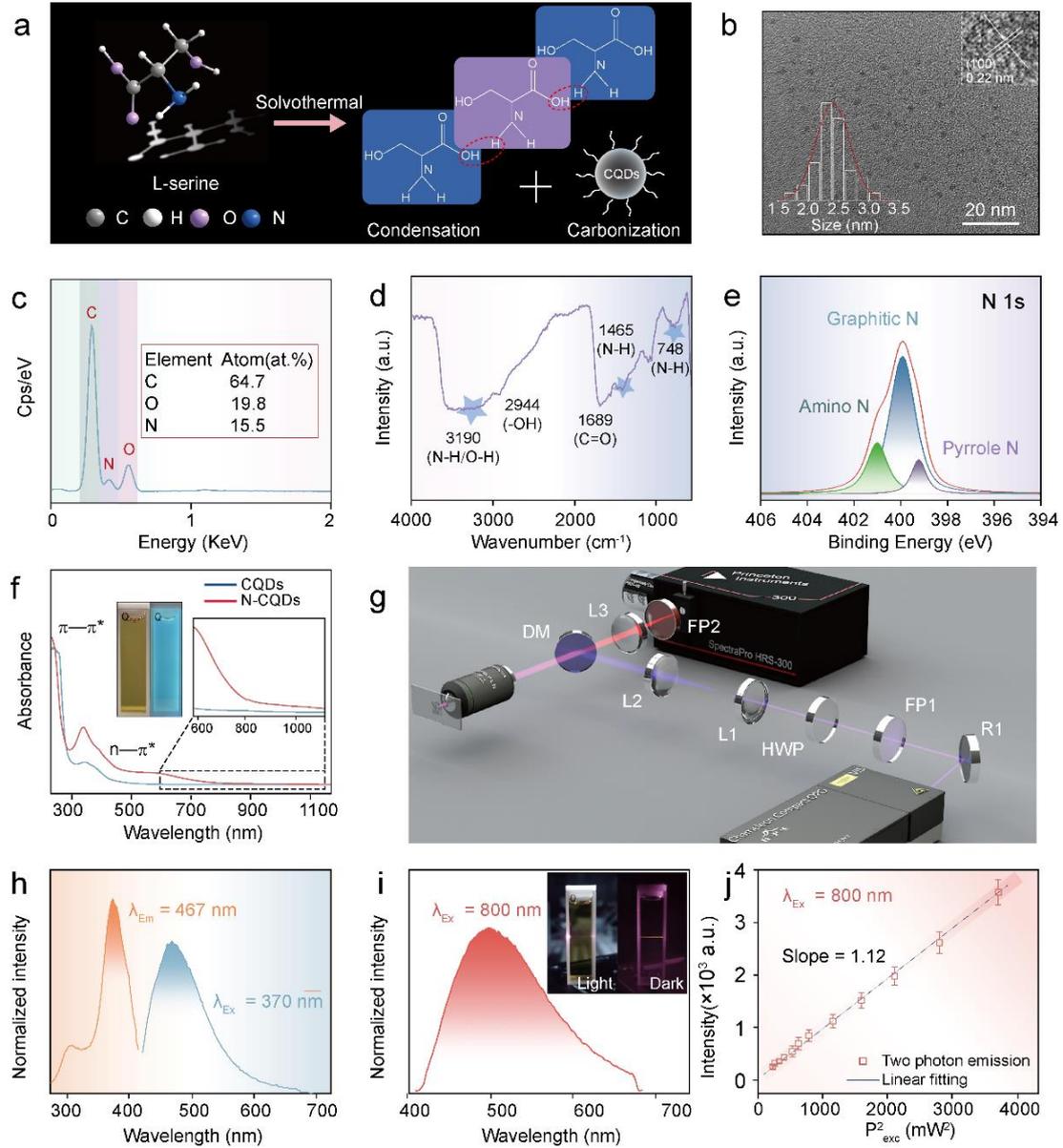

**Figure 2. Characterizations of N-CQDs. a** Synthesis scheme of N-CQDs. **b** TEM image of N-CQDs, the insets are the related high-resolution TEM and the size distribution of N-CQDs. EDS profile obtained from SEM analysis (**c**) and FTIR spectrum (**d**) of N-CQDs. **e** High-resolution XPS spectra of N 1S of N-CQDs. **f** Absorption spectra of CQDs and N-CQDs. The zoom illustration indicates the extended absorption band of N-CQDs in the near infrared region and the captured images of N-CQDs without excitation and under excitation at 365 nm. **g** Scheme of high-resolution fluorescence detection system. **h** Emission spectra of one-photon induction by Xenon lamp (467 nm excitation) and photoluminescence excitation spectrum of solution-phase N-CQDs. **i** Two-photon excitation up-conversion fluorescence of dry N-CQDs on glass substrate by femtosecond laser (800 nm excitation). Insets show the corresponding upconversion images. **j** Quadratic relationship of the observed two-photon luminescent intensity of N-CQDs with the excitation laser power at 800 nm (Pexc).

To characterize the ultrafast broadband nonlinear SSPM behavior of the N-CQDs, which is crucial for evaluating their nonlinear refractive indexes ($n_2$) and third-order susceptibilities ($\chi^{(3)}$), we first characterize the nonlinear refractive properties employing Z-scan technique with femtosecond pulse laser excitation. As expected, the closed-aperture Z-scan measurements at various wavelengths (400 nm, 532 nm, 600 nm, 740 nm, 800 nm, and 1064 nm) reveal explicit valley-peak signature patterns indicative of self-focusing effect (Figure S5). Furthermore, the large $n_2$ are determined at multiple wavelengths based on the scalar Z-scan theories (Equation S2), showing large values of ~$10^{-6}$ cm$^2$/W (Table S1), confirming the strong optical nonlinearity of the N-CQDs.

Subsequently, the temporal evolution of the SSPM diffraction rings at the corresponding wavelengths are recorded (Figure S6). It can be found that the SSPM effect contains two distinct processes: SSPM ring formation and deformation. Upon femtosecond laser pulse excitation, the diffraction rings initially expand into a circular pattern at a short time due to the optical Kerr effect, followed by the increased radius and the diffraction ring numbers at ~0.2 s. While they quickly collapse inward in the lower half of the region without change in the number of rings. All the obtained multiple diffraction ring patterns remain stable in a hemispherical shape at last at ~1 s, which can be considered as the ring deformation caused by horizontal axial gravity effect and thermal convection.[39]

All snapshots of SSPM over time can be unambiguously captured at all experimental wavelengths, indicating that the N-CQDs possesses broadband nonlinear optical response under femtosecond laser excitation. Notably, the apparent temporal resolution is limited by the intrinsic response frequency (50 Hz) of CCD detector. To address this, we analyze the intrinsic physical processes and response times of the N-CQDs to trace ultrafast relaxation processes using a time-resolved, wavelength-degenerate pump-probe experiment. Following laser excitation at 400 nm, the transient absorptions as a function of the pump-probe delay time are detected at 532 nm, 600 nm and 740 nm, respectively. The same characteristic behavior was observed for all the probe wavelengths. Specifically, the N-CQDs immediately shows a negative signal covering the range of exciton absorption and reached its maximum within a few femtoseconds after pump excitation, which indicates the photo bleaching phenomenon.[40] It can be considered that one electron-hole pair can be generated when the sample absorbs a photon of the pump light.

To obtain the timescales of those processes, the recovery kinetics of the bleaching signal are further fitted by a biexponential decay model (Table S2), in which two different time constants separately representing two photon-induced free carriers and electron–hole recombination are unraveled. The fitted curves (Figure 3a solid lines) are in excellent agreements with the experimental results. The average electron lifetimes of the N-CQDs exhibit ultrafast nonlinear response times of $\tau_1$ at three wavelengths of 520 *fs*, 750 *fs* and 6.84 *ps*, respectively, which are close to the results previously reported for quantum well and quantum dots.[41, 42] These extremely fast recombination times may stem from the interband and intersubband transitions influenced by quantum-confinement effects.[43] The ultrafast response of the N-CQDs further highlights their strong potential for application in next-generation nonlinear and ultrafast photonic devices.

Next, we further verify large and tunable optical nonlinearities of the N-CQDs based on the SSPM method. Typical SSPM patterns formed at six different wavelengths are shown in Figure 3b. It is easy to notice that the SSPM response shows a proportionally linear relation between the diffraction ring number *N* and laser intensity *I* in all wavelengths, which is the critical characteristic of the SSPM nature of the patterns owing to the Kerr nonlinearity.[44] For the same *N*, the required *I* decreases with the increase of wavelength. The illustration shows a part of the SSPM images obtained from a CCD at different wavelengths. Through the vertical light geometry, the results present a perfect circle, which confirms the ring distortion caused by heat convection.

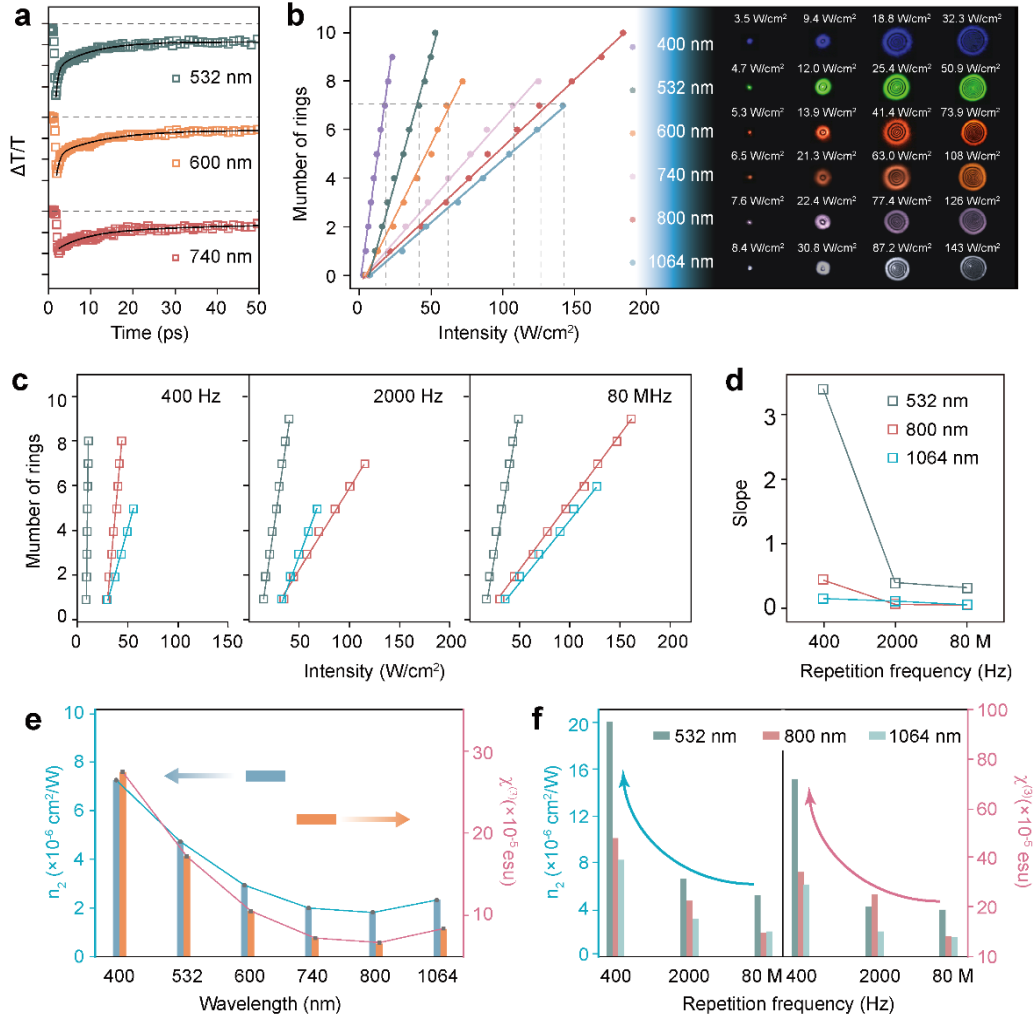

**Figure 3. Ultrafast, broadband and large nonlinear optical measurements of N-CQDs. a** Transient absorption spectroscopy of the N-CQDs with three different wavelengths. Solid lines show the fitted curves using a bi-exponential decay model. **b** Intensity dependence of the diffraction ring number (left) and snapshots of the self-diffraction ring patterns (right) at different wavelengths from 400 ~ 1064 nm. Solid lines are guides for the eye for the overall linear relation. **c** Variation of the diffraction ring numbers to the incident intensity via modulate frequencies of $f$ = 400 Hz, $f$ = 2000 Hz, and $f$ = 80 M Hz. **d** Trend of the slope determined by the ratio between the number of rings and intensity at different repetition frequencies. **e** Measured nonlinear refractive indexes and nonlinear susceptibilities of the N-CQDs at different wavelength. **f** Nonlinear refractive indexes and nonlinear susceptibilities of the N-CQDs at three repetition frequency. The arrow indicates the increased direction of the nonlinear coefficients.

To further investigate frequency-dependent modulation effects, we perform additional experiments using a mechanical chopper to vary the laser repetition frequency. Figure 3c illustrates the relationship between $N$ and $I$ under different repetition frequencies at three wavelengths (532 nm, 800 nm and 1064 nm). The results show that shorter wavelengths require lower excitation energy to achieve the same ring number. For example, when $N$=8 under 532 nm excitation, the corresponding $I$ at modulator frequencies of $f = 80$ MHz, $f = 2000$ Hz, and $f = 400$ Hz are 38.4 W/cm$^2$, 37.6 W/cm$^2$ and 10.6 W/cm$^2$, respectively, revealing a fourfold difference. The results indicate that the diffraction ring pattern is strongly influenced by modulation frequency, which might be attributed to a negative correlation with thermal lensing effect.[45] When the frequency is modulated at 400 Hz and 2000 Hz, there is no significant change at 1064 nm wavelength because it is not affected by the chopper. This relationship is given in Figure S7, where the intensity remains almost constant when the repetition frequency changes from 400 to 3000 Hz. To quantify the contribution of repetition frequency, the slope ($N/I$) is analyzed at different excitation wavelengths (Figure 3d). The effect of frequency modulation is strongly dependent on the wavelength of the incident beam, which can be seen in the wavelength of 532 nm with the largest $N/I$ value. Meanwhile, the $N/I$ is directly related to the calculation of $n_2$, which can also imply a large nonlinear effect at short wavelength incidence. These results support the conclusion that both wavelength and modulation frequency strongly affect the nonlinear optical behavior of the N-CQDs.

On the basis of the above results, $n_2$ and ($\chi^{(3)}$) of the N-CQDs can be deduced through the analysis of the SSPM diffraction ring patterns (Equation S3). As shown in Figure 3e, under unmodulated femtosecond pulse laser excitation, the values of $n_2$ and $\chi^{(3)}$ are found to be on the order of ~$10^{-6}$ cm$^2$/W and ~$10^{-5}$ e.s.u., respectively, which are consistent with the results derived by the Z-scan method. It is also found that the obtained $n_2$ and $\chi^{(3)}$ are higher at the short wavelength excitations. For instance, $n_2$ and $\chi^{(3)}$ at 400 nm and 1064 nm in turn are $8.4 \times 10^{-6}$ cm$^2$/W ($3.26 \times 10^{-4}$ (e.s.u.)) and $2.0 \times 10^{-6}$ cm$^2$/W ($7.7 \times 10^{-5}$ (e.s.u.)), respectively.

Additionally, we further study the variation of the nonlinear optical coefficients at different repetition frequency shown in Figure 3f. The related maximum nonlinear response parameters appeared under the excitation of 400 Hz. They are, respectively, $2.0 \times 10^{-5}$ cm$^2$/W ($7.2 \times 10^{-4}$ (e.s.u.)), $1.0 \times 10^{-5}$ cm$^2$/W ($3.4 \times 10^{-4}$ (e.s.u.)) and $0.8 \times 10^{-5}$ cm$^2$/W ($2.9 \times 10^{-4}$ (e.s.u.)), enhancing 3.8 (3.9) times at 532 nm, 5 (4.2) times at 800 nm and 3.9 (3.7) times at 1064 nm comparing to that excited at 80 MHz. For clarity, the nonlinear parameters of the N-CQDs at different wavelengths and repetition frequencies are calculated and summarized in Table S3. Overall, it can be clearly seen that the ultrafast nonlinear optical responses can be improved by both wavelength and frequency modulation. Remarkably, by comparing these reported results in Table S4, the optical nonlinearities of the N-CQDs have considerable advantages over all previously reported nonlinear materials, indicating the great potential for applications in nonlinear optical and all-optical modulation devices.

An essential observation is that any excitation wavelength other than 400 nm is insufficient to induce SSPM in the CQDs (Figure S8), thus it is imperative to elucidate the broadband SSPM mechanism of the N-CQDs. Unlike most 2D materials, where the single-photon absorption dominates in the visible and near-infrared range, a hybrid single-photon and two-photon SSPM mechanism is proposed for the N-CQDs (Figure S9). Structurally, the CQDs consist of graphene nanofragments with evident graphene lattices and edge-bound chemical groups. Upon nitrogen doping, electron-donating N atoms introduce hydrophilic surface functional groups, such as -NH$_2$, -OH, and O=C-OH, effectively extending the conjugated sp$^2$ domains compared to pristine CQDs.[32, 46] This structural evolution results in a wide optical absorption and an increased TPA cross section. On the other hand, the absorption-derived bandgap (Figure S10) also exhibits that the band-gap of the N-CQDs (2.25 eV) is lower than that of the CQDs (2.54 eV), facilitating TPA at sub-bandgap and near-bandgap excitation wavelengths. Therefore, depending on the above electronic structure changes, the TPA-induced SSPM together with one-photon-induced SSPM can occur despite excitation light below or near the band gap, enabling broadband SSPM across the visible and near-infrared regions.

To leverage these merits of the N-CQDs, we experimentally confirm broadband ultrafast AOS application with low excitation threshold based on the SSPM for the first time (Figure 4a and 4b). In experimental configuration, two femtosecond pulse beams are used as a control and signal lights, respectively (Figure S11). The control light is employed to modulate the propagation of weak signal light, resulting in the SSPM rings increase via the enhancement of the incident intensity. Upon 532 nm excitation, ultrafast broadband AOS is demonstrated at different signal wavelengths (400 nm, 740 nm, 800 nm and 840 nm). When the intensities of both beams are simultaneously lower than the threshold value for exciting the SSPM effect, the result remains the original far-field spots, defined as the "Off" state. Only if the input intensity of the control light increases and the two-beam coherence occurs, the double self-diffraction rings appear, defining the "On" state. Comparably, the multi-colored SSPM is also observed under excitation of 600 nm, which enables ultrafast multi-channel AOS. In addition, with excitation of 532 continuous light, the broadband AOS at 400 nm, 532 nm, 800 nm and 1064 nm are also demonstrated (Figure S12). By exploiting multi-colored manipulation, this fascinating performance here makes it possible for widespread applications requiring broad bandwidth.

Figure S13 further quantifies the modulation dynamics by plotting the modulation relationship between the control and signal lights for the N-CQDs with signal light intensity fixed, respectively. As the control light intensity (532 nm, 600 nm) increases, the ring number of signal light (400 nm, 740 nm, 800 nm and 840 nm) increases linearly. This linear enhancement is independent of the signal light intensity, indicating that the control light intensity can effectively modulate the nonlinear phase transition of the signal light. Notably, if the initial intensity of the signal light is below the excitation threshold, the intensity of the control light needs to fill this gap and increase the energy to achieve the "On" state.

The dual-beam SSPM-derived *N/I* slopes (Figure 4c) at different signal and control wavelengths further confirm this tunable nonlinearity. There is a higher slope (nonlinearity) at 532 nm excitation rather than 600 nm excitation. Consequently, the slope decreases from 0.25 (at 400 nm) to 0.08 (at 1064 nm) under excitation of 532 nm control light. The inset quantifies that for a 400 nm single beam at 2.54 $W/cm^2$, a minimum excitation threshold of only 2.2 $W/cm^2$ at 532 nm is required to induce a phase π change between the adjacent rings of the SSPM.

Further, a strategy for constructing an ultrafast AOS process based on the nonlinear optical response of N-CQDs is schematically illustrated in Figure 4d. The dual-femtosecond pulse laser system (400 nm and 532 nm), as depicted in Figure 4c, serves as the control and signal light sources, respectively. When the two synchronized femtosecond beams interact with the N-CQDs, a periodic state of rapid excitation and relaxation is generated within the electronic structure. For a given period, the population of excited-state carriers governs the output signal intensity, thereby creating a temporal window for demonstrating the SSPM effect. This high-level light detection state represents the "ON" state of the ultrafast AOS, with an anticipated timescale in the order of hundreds of femtoseconds. The persistence time of this continuous rising level in the ON state is defined as the operational duration of the AOS. Conversely, the excited-state carriers relax toward the ground state, which results in a depletion of the photoexcited carrier population. Thus, the nonlinear response of N-CQDs is completely suppressed in this process, which represents the "OFF" state. In other words, the turn-off duration of the AOS can be ultimately quantified by the intrinsic carrier relaxation dynamics of the N-CQDs.

To the best of our knowledge, this study represents the first demonstration of a dual-femtosecond beam-controlled ultrafast all-optical switch (AOS) utilizing N-CQDs with exceptional switching speed. While previous AOS implementations employing continuous Gaussian laser beams have been demonstrated in materials such as black phosphorus, MXene, and tin sulfide, these systems failed to precisely define the switching speed characteristics. regardless of whether the continuous beam serves as the signal or control beam, it results in persistent occupation of the excited state (as illustrated in Figure

4e(1-3)), effectively maintaining the system in a constant operational state. In contrast, our implementation of dual femtosecond beams (shown in Figure 4e(4)) enables distinct on/off switching behavior, as evidenced by the alternating output response. The N-CQDs-based ultrafast AOS demonstrates remarkable response characteristics, with measured rise and recovery times of ~520 *fs* and ~8.54 ps, respectively. These values represent nearly an order-of-magnitude improvement compared to state-of-the-art materials, including Silicon and ITO-based optical switches.[20, 47] Meanwhile, the effective nonlinear figure of merit (FOM) of the N-CQDs—calculated from broadband femtosecond Z-scan measurements (Figure S14 and Table S5)—reaches values as high as 10 across a broad excitation wavelength range from 400 to 1064 nm (Figure 4f). Notably, the FOM of N-CQDs ranks among the highest reported to date for nonlinear nanomaterials over such a wide spectral band, as compared to state-of-the-art counterparts (Figure 4f and Table S6). Collectively, the N-CQDs-based AOS achieves an all-in-one combination of broad operational bandwidth, ultralow switching threshold, and sub-picosecond-scale response speed—three critical performance metrics seldom realized together in the N-CQDs ( shown in Table 1). This exceptional performance underscores the potential of N-CQDs for efficient and ultrafast nonlinear light manipulation with high bandwidth and parallelism. Consequently, this breakthrough in realizing ultrafast, low-threshold, and broadband AOS capability opens new avenues for advanced nonlinear photonics applications.

**Table 1.** Comparison of three indicators of AOS.

| Materials | Wavelength | Time | Intensity | Ref. |
|---|---|---|---|---|
| TiN | 1550 nm | 763 *fs* | 17 MWcm$^{-2}$ | [48] |
| Silicon | 1550 nm | 14.8 *ps* | / | [47] |
| CsPbBr$_3$ | 400 nm | / | 32 MWcm$^{-2}$ | [49] |
| Al:ZnO | 778 nm | / | 33.3 GWcm$^{-2}$ | [43] |
| In:CdO | 2080 nm | 800 *fs* | / | [19] |
| PhNoC | 1550 nm | 18 *ps* | / | [50] |
| tellurium | 1550 nm | 276.3 *µs* | / | [6] |
| Ag:(PCBM:P3HT) | 800 nm | 33.6 *ps* | 70 kWcm$^{-2}$ | [51] |
| Corrugated ITO | 1450 nm | 350 *fs* | 308.5 GWcm$^{-2}$ | [20] |
| C153:polystyrene | 788 nm | / | 110 kWcm$^{-2}$ | [26] |
| **N-CQDs** | **400 ~ 840 nm** | **520 *fs*** | **2.2 Wcm$^{-2}$** | **This work** |

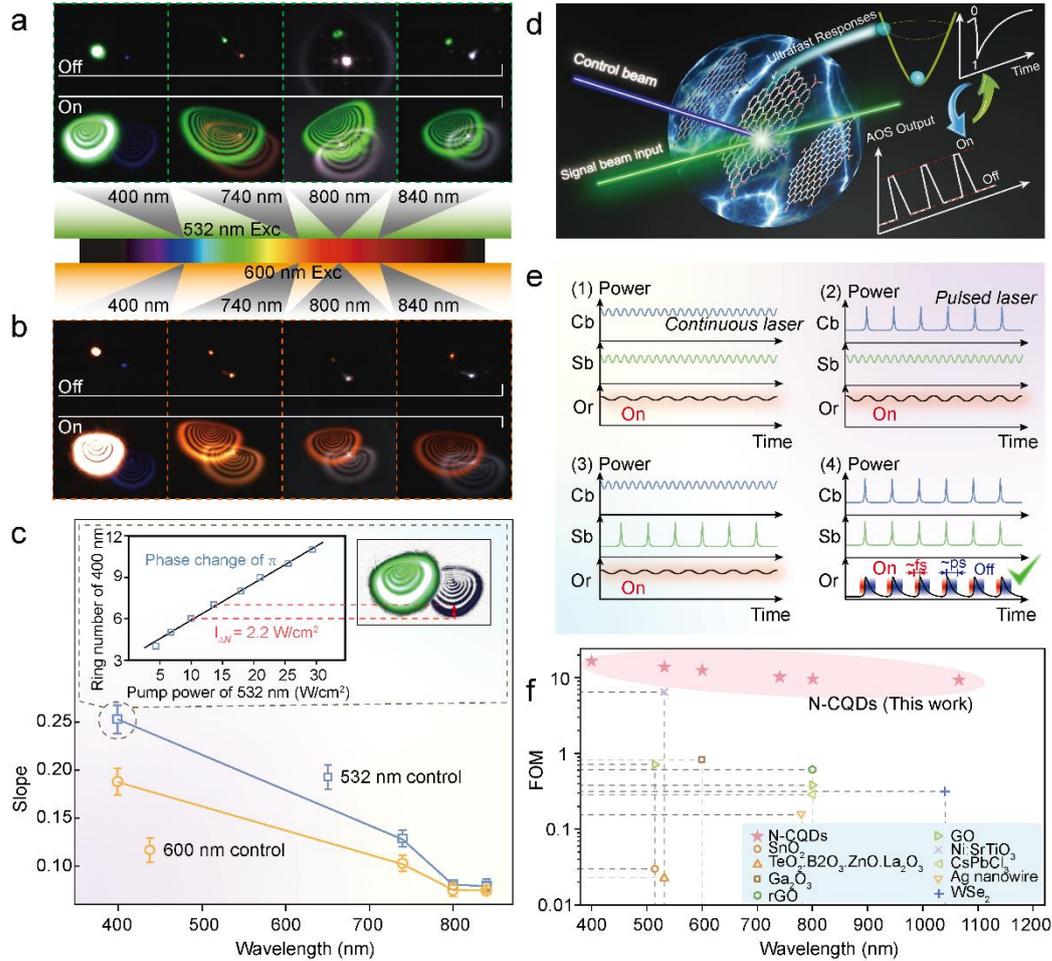

**Figure 4. Schematic of ultrafast low-threshold and broadband AOS based on SSPM of the N-CQDs.** Actual images obtained using femtosecond pump light (**a**) ($\lambda$ = 532 nm) and (**b**) ($\lambda$ = 600 nm) to modulate the femtosecond probe light ($\lambda$ = 740, 800, 840 nm) in the N-CQDs-based AOS demonstration. **c** Slope determined by the ratio between the number of signal ring and pump intensity. The inset is the lowest threshold curve using 532 nm laser beam to control 400 nm laser beam. **d** Schematic illustration on the operating principle of the ultrafast AOS based on the nonlinear optical effect. **e** Demonstration of all-optical switch dependent on incident laser. Cb: Control beam; Sb: Signal beam; Or: Optical response of N-CQDs. **f** Comparison of FOM values between N-CQDs and typical reported nonlinear optical materials.

## 4. Conclusions

In this work, we have successfully developed a novel dual-beam femtosecond light-control-light system to realize ultrafast low-threshold and high-efficiency AOS, enabled by the broadband ultrafast SSPM of N-CQDs. We first precisely engineer heteroatom-hybridized CQDs via tailored amino-functionalization, incorporating an optimized donor-acceptor configuration. This design enhances the transition probability and strengthens the n-$\pi$* absorption band, yielding ultrafast response characteristic, high electron transport efficiency and wide-band absorption property. The synthesized N-CQDs, prepared through a facile hydrothermal synthesis process demonstrate exceptionally rapid carrier dynamics at *fs* to *ps* scale through the transient absorbance spectroscopy, which endows a prerequisite for achieving ultrafast AOS. Subsequently, we systematically investigate the dependence of the dependence of SSPM-enabled optical nonlinearity on laser intensity, incident wavelength and repetition frequency, from which we observe that the number of the diffraction rings is closely proportional to the former two parameters, whereas exhibits an opposite trend relative to the latter one. Concretely, at 532 nm and a repetition frequency of 400 Hz, we obtain a sharply escalated slope value (N/I), which in turn enables the derivation of the corresponding optical nonlinear parameters. In this case, $n_2$ and $\chi^{(3)}$ of the N-CQDs are determined to be approximately $10^{-5}$ cm$^2$/W and $10^{-4}$ esu, which are at least two orders of magnitude higher than the state-of-the-art, thus making the high-efficiency AOS possible.

Furthermore, our proposed femtosecond dual-beam light-control-light system is shown to feature ultra-low excitation threshold of 2.2 W/cm$^2$, three orders of magnitude lower than the state-of-the-art, by resorting to ultrafast multi-wavelength SSPM, which is impossible to achieve for the carbon nanotubes reported previously. This strategy opens the door to multi-channel and energy-saving AOS. Ultrafast fluorescence spectroscopy further confirms that N-CQDs exhibit broader absorption band and larger TPA cross sections than the pristine CQDs, which is attributed to the synergistic interaction between the nitrogen-derived structure and the nitrogen or oxygen-related surface states. Such an exotic effect challenge conventional understanding by demonstrating that SSPM can be induced in materials even below their bandgap, thereby unlocking a roadmap for the two-photon and defect-assisted SSPM and facilitating the in-depth understanding on the physical origin of ultrafast low-threshold and high-efficiency AOS.

Looking forward, we anticipate that other alternative 0D quantum dot species, such as graphene quantum dots, carbon nanodots and polymer dots, can be synthesized in a similar manner to induce a superior SSPM, which will bring endless expectation for future AOS-dependent photonics applications. We believe that our findings provide crucial insights for designing novel next-generation 0D carbon quantum nonlinear materials, and open up a compelling roadmap based on the N-CQDs for developing ultrafast multifunctional AOS.

## 5. Material Design, Synthesis and Characterization

5.1 Computational methods

The geometry structures were optimized using the Gaussian 16 program package. The M06-2X hybrid functional was used during the geometry optimization. The Def2-SVP basis set was used during the whole calculation The solvent effect of water was considered, using the polarizable continuum model (PCM) during geometry optimization and Solvation Model Density (SMD) implicit solvent model during single point calculations.

All single point calculations were carried out with ORCA 6.0.0 program package. the range-separated hybrid CAM-B3LYP with Van der Waals interaction (DFT-D3 method with Becke-Jonson damping) was used with a Def2-SVP basis set during the UV vis spectrum calculation. The HOMO (Highest occupied molecular orbital) and the LUMO (Lowest unoccupied molecular orbital) was extracted from a single point calculations of B3LYP with Van der Waals interaction (DFT-D3 method with Becke-Jonson damping) and Def2-SVP basis set. All post calculation analysis were carried out on Multiwfn 3.8 Dev and visualized by VMD 1.9.4.

5.2 Synthesis of N-CQDs

For synthesis of N-CQDs, 1.31 g of L- serine were ultrasonically (20 min) dissolved in 36 mL acetic acid (2 wt%) in a 100 mL beaker to form a transparent solution. Then, the mixed solution was transferred into a Teflon autoclave for solvothermal reaction. After heating at 200 ℃ for 12 h, the autoclave was cooled down to room temperature naturally. The obtained solution was treated with ultrasound for 10 min and further dialyzed against deionized water with a dialysis tube (3500 Da) for 24 h to remove unreacted small molecular species. After that, the dialysis solution was filtered by 0.22 μm hydrophilic filters, and the purified N-CQDs solution was dried under vacuum freeze-dryer. Finally, the purified N-CQDs powder was collected for further characterization and analysis. By the way, the CQDs samples were synthesized by the same procedure using citric acid monohydrate as raw material.

5.3 Materials characterization

The crystalline structure of the N-CQDs powder was measured via X-ray diffractometer (Shimazu, Japan) with Cu Kα radiation (λ = 0.15406 nm). Energy Dispersive Spectrometer (EDS) measure result was obtained by field emission scanning electron microscope (JSM-IT700HR). Fourier transform infrared (FT-IR) spectra were collected via a TENSOR27 Fourier transform infrared spectrometer. A Raman microscopy system (Renishaw Invia Qontor) was used to analyze the degree of structural order. The excitation of Raman scattering was performed with the solid state 785 nm laser. The size and surface topography of the as-prepared N-CQDs were noted by transmission electron microscope (TECNAI G2

F20). Ultraviolet-visible (UV–vis) transmittance spectra of the films were acquired on a UV–vis spectrophotometer (Carry-5000 Varian).

## 6. Broadband ultrafast SSPM and AOS configuration

Figure 1a shows the schematic of the experimental setup, which builds up between the parallel and vertical excitation geometry patterns by means of single wavelength pulse laser and the configuration of the light-control-light system making use of dual wavelength beam excitation. In the given experiment device, a ultrafast fiber laser (Chameleon Ti: Sapphire Lasers) with 680~1080 nm, 80 MHz repetition rate and ~100 $fs$ pulse width is employed in the experiment. Along with the Chameleon pump laser (Chameleon Compact OPO-Vis), the wavelength of the output laser can be extended from the visible through to the near infrared. At the same time, an optical chopper system (MC2000B/Thorlab) placed in the optical path is used to tune the output repetition rate of the laser pulse. In the propagation process, a laser beam is focused onto the N-CQDs solution by a focusing lens (focal length $f$ = 150 mm). In the plane of observation, a CCD (DCC1645C/Thorlab) placed ~30 mm away from the cuvette, which can respond from the spectral ranges of 350-1100 nm, is used to detect the entire process of the formation of far-field self-diffraction intensity pattern. At each SSPM stage occurs, the optical power value is recorded by the optical power meter (PM100D/Thorlab), which has a power probe that can measure the wavelength range of 200 nm ~ 1100 nm. It should be noted that in the process of vertical measurement of SSPM. In order to avoid the influence of gravity, the laser beam is incident strictly along the surface of the cuvette in a vertical direction.

**Acknowledgment**

The authors would like to acknowledge financial support from the National Natural Science Foundation of China (Grants Nos. 12574333, 62575305, 62305275, and 11974258); Innovation Science Fund of National University of Defense Technology (Grant No. 25-ZZCX-JDZ-28); Youth Independent Innovation Science Foundation of the National University of Defense Technology (Grant No. ZK24-20).